\documentstyle[11pt,newpasp,psfig]{article}

\begin{document}

\title{Star Formation in Clusters: Subclustering, Cloud Fragmentation and the
Origin of the Stellar IMF}
\author{Leonardo Testi}
\affil{Osservatorio Astrofisico di Arcetri, Largo E. Fermi 5, I-50125 Firenze,
       Italy}

\begin{abstract}
We review recent high spatial resolution millimeter continuum and spectral line 
observations of (proto-)cluster regions. These observations reveal that the 
mass distribution of prestellar cores is consistent with the initial mass 
function for field stars suggesting that the IMF is connected to the molecular
clouds structure or the cloud fragmentation processes, rather than the 
details of the star formation mechanism. We discuss the evidence for at-birth 
subclustering independently obtained for two separate protoclusters.
The presence of subclustering within coherent subclumps suggests that
the fragmentation process is hierarchical and that young stellar clusters are 
assembled by independent substructures. The local (proto-)stellar density
and star formation efficiency within the subclusters are much higher
than the average values for the entire star forming cloud.
Unfortunately, current observations are sparse and limited to the nearest 
star forming regions, for spatial resolution and sensitivity considerations,
the future millimeter wave arrays, especially ALMA, will allow to expand
considerably the sample of observed regions.
\end{abstract}


\section{Introduction}

There is now a fairly robust theory of how isolated, low-mass stars form
(e.g. Shu et al.~1987; 1993). One notable omission, is the failure
to predict the resulting stellar masses. Stars in the field are known
to be distributed according to a well defined mass function
(e.g. Salpeter~1955; Kroupa et al.~1993),
and a complete theory of star formation should explain
this mass distribution, the Initial Mass Function (IMF). Since it now
seems very likely that most stars form in clusters rather
than in isolation (Clarke et al.~2000),
the way in which the IMF originates must be closely
related to the way stars form in clusters.
Indeed, young embedded clusters appear to be populated by stars with a 
mass distribution very close to that observed in the solar neighbourhood
(Hillenbrand~1997).

The different theories of the IMF have been recently reviewed by several
authors (e.g. Elmegreen~2001), two different phylosophies are generally
considered: i) the observed distribution of stellar masses may result
naturally from the protostellar accretion process, such as due to feedback
effects (e.g. Fatuzzo \& Adams~1996) or competitive accretion
(Bonnell et al.~1997), alternatively, ii) it may be related to the
molecular clouds structure and fragmentation processes 
(Elmegreen~2000; Myers~2000; Padoan \& Nordlund~2001).
If the fragmentation hypothesis
is correct, the mass distribution of prestellar condensations in
cluster-forming molecular cores should follow the IMF of the stars in
young embedded clusters and in the field.
Millimeter wavelength interferometry offers both a high spatial
resolution and high sensitivity that are critical to shredding light
on this problem. At 3~mm the thermal emission from dust in the condensations is
probably optically thin, allowing a reasonable estimate of clump 
masses, and thus an insight into the clump mass spectrum.
High resolution imaging of extended areas of the sky at millimeter 
wavelengths can be performed by means of the multi-field interferometric
imaging technique (see Testi \& Sargent~2000 for its implementation at
the OVRO array).
In addition, an interferometer provides simultaneous observations of molecular
line and broad band continuum emission. Any contamination of the continuum
flux by molecular line radiation can therefore be eliminated. Finally,
thanks to the interferometric filtering capability, smooth, extended emission
from the molecular cloud in which cores are embedded is resolved out.

Using the Owens Valley Radio Observatory millimeter wave array,
we have begun a program of high resolution,
millimeter-wave mapping of molecular cloud cores with a view to
establishing whether the prestellar clump mass function and the IMF are
in general similar.
Here we review our results for the Serpens star-forming core
(Testi \& Sargent~1998, Testi et al.~2000). The results of other surveys 
towards other regions as well as the promise of the next generation
of millimeter arrays (ALMA) are also discussed.

\section{The Serpens core}

At a distance of 310~pc (de~Lara et al.~1991), and with an angular extent
of few arcmin (Loren etal.~1979), the Serpens molecular core
is an ideal target to search for
compact prestellar and protostellar condensations.
Inside the 500-1500~M$_\odot$ core is a young stellar cluster of approximate
mass 15-40$\rm\,\,M_\odot$ (Giovannetti et al.~1998), while far-infrared
and submillimeter observations reveal the presence of a new generation of
embedded objects (Casali et al.~1993; Hurt \& Barsony~1996).

At low spatial resolution ($\sim50$\arcsec) the Serpens core
appears as a massive rotating molecular clump which is fragmented
in two main sub-clumps with size-sclae of $\sim$0.2~pc,
each with a remarkable internal velocity coherence (Testi et al.~2000;
Olmi et al.~2001). Our millimeter interferometric observations in the 
99~GHz continuum and CS(2--1) line cover the inner regions of the 
clump and encompass all the sub-clumps structures seen at low
spatial resolution. The details of the observations are given elsewhere 
and will not be reviewed here (see Testi \& Sargent~1998; 2000).

\begin{figure}
\centerline{\psfig{figure=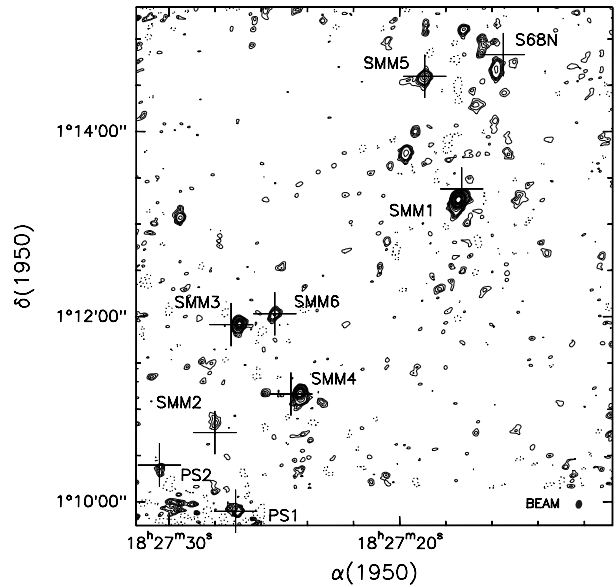,height=7.9cm,angle=-90}}
\caption[]{The 3~mm continuum hybrid mosaic of the Serpens core (Testi \&
Sargent~1998).
The rms in the map is $\sim 0.9$~mJy/beam, and the synthesised 
beam, indicated by a black ellipse in the lower right corner, is
$5.5^{\prime\prime}\times 4.3^{\prime\prime}$ FWHM. Submillimeter and
far-infrared sources from Casali et al.~(1993) and Hurt \& Barsony~(1996)
are indicated.}
\label{fcontmos}
\end{figure}

\begin{figure}
\centerline{\psfig{figure=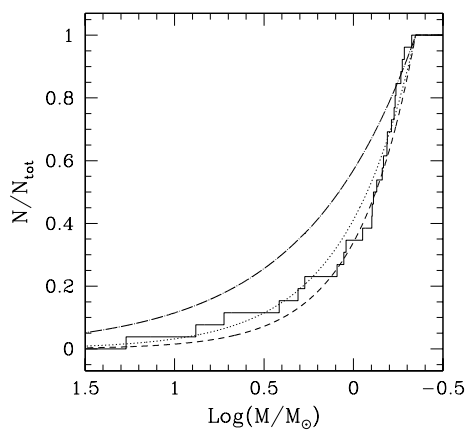,width=6.5cm}}
\caption[]{Cumulative mass spectrum of the protostellar and 
prestellar cores found in the 3~mm continuum mosaic shown
in Figure~\ref{fcontmos}. The dotted line is the best fitting power law
$\rm dN/dM\sim M^{-2.1}$, the dashed line shows 
the Salpeter IMF, and the dot-dashed line depicts the 
power law typical of gaseous clumps in molecular clouds,
rejected by the Kolmogorov-Smirnov test at the 98\% confidence level (Testi \&
Sargent~1998).}
\label{fmcum}
\end{figure}

The 3~mm continuum high resolution mosaic (Fig~\ref{fcontmos}) further
resolve the structure of the clump in a large number ($\sim$30) of cores.
The mass of each of the dust condensations in Figure~\ref{fcontmos}
can be calculated, assuming reasonable values for the parameters of the
emitting dust (see Testi \& Sargent~1998 for details).
Using existing near- and far-infrared observations (Givannetti et al.~1998;
Hurt \& Barsony~1996), the few condensations associated with already
formed stars were eliminated from our list of candidate prestellar and
protostellar candidates.
Figure~\ref{fmcum} displays the cumulative mass spectrum for the remaining 
26 continuum sources above a the $\sim$4.5$\sigma$ peak threshold of
4~mJy/beam. This mass distribution is well fitted by 
a power law $\rm dN/dM\sim M^{-2.1}$, significantly 
steeper than the mass spectrum of larger-scale gaseous clumps in molecular
clouds $\rm dN/dM\sim M^{-1.7}$ (Williams et al.~2000), which
is rejected by the Kolmogorov-Smirnov test at the 98\% confidence level.
Our power law exponent, $-$2.1, is close to the Salpeter~(1955)
IMF value, $-$2.35, and very similar to that suggested recently by 
Kroupa et al.~(1993, see also his review in this volume) for solar and
slightly subsolar stellar masses.
The inferred masses and sizes of the cores suggest that these
are self gravitating and are going to be the progenitors of
individual stellar systems (single or multiple stars).
Our result are supported by the findings by Motte et al.~(1998), their IRAM-30m 
1.3~mm continuum maps of $\rho$-Ophiuchi, another cluster forming core,
show that the mass spectrum of the prestellar and protostellar clumps
within the core appear consistent with the IMF.
Taken together these observations provide compelling evidence for 
a ``clump fragmentation'' origin for the stellar IMF.


\section{Hierarchical fragmentation and sub-clustering}

\begin{figure}
\centerline{\psfig{figure=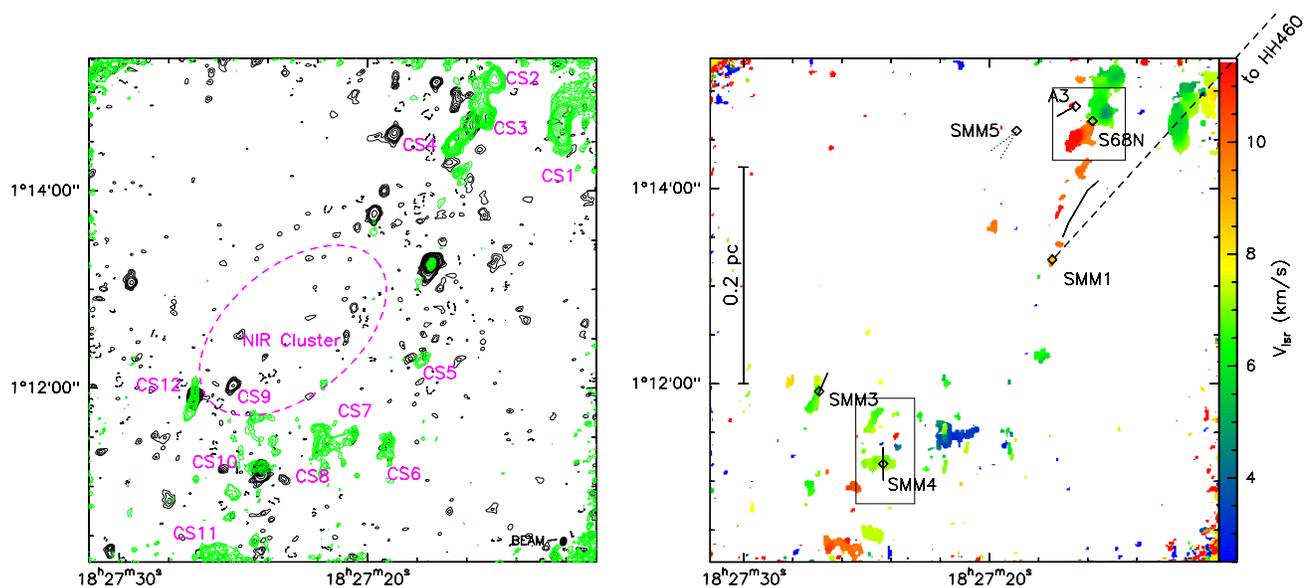,height=7.9cm}}
\caption[]{a) The OVRO CS(2--1) integrated intensity contour map at
$5^{\prime\prime}\!.5\times 4^{\prime\prime}\!.3$ resolution, superposed on the 
millimeter continuum mosaic. b) CS(2--1) first moment map, the direction of 
optical and near infrared outflows are marked (adapted from Testi et al.~2000).
}
\label{fcs}
\end{figure}

In Figure~\ref{fcs} we show the OVRO CS(2--1) integrated intensity 
and first moment mosaics for the Serpens core.
Most of the extended emission seen in the single dish maps (McMullin et
al.~1994; 2000; Olmi et al.~2001)
is missing, probably because the optically thick, extended,
molecular line emission is resolved out by the interferometer.
Most of the CS(2--1) emission we detect appears to be due to the
known outflow sources. This interpretation is confirmed by the coincidence
of the CS features with emission 
of molecular tracers enhanced in outflow regions.
In Figure~\ref{fcs} we indicate these outflows along with their approximate 
orientations (Testi et al.~2000; Olmi et al.~2001; but see also Wolf-Chase et al.~1998; Hogerheijde et al.~1999).

We find that the millimeter continuum cores are mostly confined within the two 
subclumps detected in the large scale molecular line emission. Moreover,
the direction of the flows 
from the protostars within a single subclump
are well aligned (Figure~\ref{fcs}).
These findings together with the internal velocity coherence
of the subclumps suggest a hierarchical picture for the fragmentation of the molecular cloud, in which coherent velocity structures (the sub-clumps) fragment
to form the millimeter continuum cores, which are the progenitors of
single stellar systems. Thus, within each of the
sub-clumps an independent sub-cluster of protostars is being assembled.
A similar behaviour is also observed in the $\rho$-Oph cluster, where
Johnstone et al.~(2001) noted a sub-clustering of the prestellar cores 
over the same scale length observed in Serpens, $\sim0.2$~pc. It is 
interesting to note that this scale-length is of the same order as the
scale-length of the externally
driven shear flow turbulence in non star-forming clouds, i.e. clouds
without an internal perturbance (LaRosa et al.~1999). It is, however, 
premature at this point to argue if and how this process may affect
star formation in molecular clouds.

The picture that comes out
is that of a stellar cluster being assembled in denser subclusters that
eventually 
will merge during the dynamical evolution of the (proto-)cluster on a timescale
of a few million years (Testi et al.~2000; Clarke et al.~2000). 
An important consequence of this view is that even though the {\it average}
(proto-)stellar density of the (proto-)cluster and the {\it total} star 
formation efficiency of the parent cloud are both rather low, the {\it local}
density and star formation efficiency are expected to be much higher.
In the case of the Serpens, while the average (proto-)stellar density 
and star fromation efficiency are $\sim$400~$\star$/pc$^3$ and
$\le 2-5$\%, both are a factor of $\sim$10 higher in the sub-clusters
(Testi et al.~2000).

The conclusions, although consistently supported by independent observations of
two different regions,
are, however, preliminary, since the results of the surveys may be 
vitiated by small number statistics. Combined millimeter and 
infrared surveys of a larger number of cluster forming cores are necessary to
ensure the required statistical significance. We note that in this respect the
next generation millimeter arrays will be the ideal instruments to attain 
the required resolution and sensitivity. In fact, at the low resolution
attainable with single dish instruments, it is not possible to resolve the
parent cores of each individual star, and the mass distribution
of the large scale gaseous clumps in molecular clouds is recovered
(e.g. Tothill \& White~2000).
A tremendous advance is expected from the next generation
millimeter array (ALMA). At the sensitivity, frequency coverage
and spatial resolution of the ALMA baseline array, it will be possible to 
map at the same linear resolution and mass sensitivity than the Serpens OVRO
mosaic
the giant star forming regions (e.g. 30~Doradus) in the Large Magellanic Cloud,
it will thus be possible to probe the initial conditions for star formation
beyond our own galaxy.

\acknowledgments

The Owens Valley millimeter-wave array is supported by NSF grant AST-96-13717.
Research on the formation of young stars and planets
is also supported by the {\it Norris Planetary Origins Project}.
Funding from the C.N.R.--N.A.T.O. Advanced Fellowship program and from
NASA's {\it Origins of Solar Systems} program (through grant
NAGW--4030) is also gratefully acknowledged.

\end{document}